\documentclass[12pt]{article}
\def\lae{\;^{<}_{\sim} \;} 

\title{\textbf{Type II Seesaw Higgs Triplet as the Inflaton for Chaotic Inflation and Leptogenesis}}
{\author{\\[1cm]
{\sc \large Chian-Shu Chen$^{1,\dag}$ and Chia-Min Lin$^{2,\ddag}$}\\
{\sl\small $^\dag$Physics Division, National Center for Theoretical Sciences, Hsinchu, Taiwan 300}\\
{\sl\small $^\ddag$Department of Physics, National Tsing Hua University, Hsinchu, Taiwan 300 }\\
{\sl\small $^\dag$ Institute of Physics, Academia Sinica, Taipei, Taiwan 115}\\
}}

\usepackage[margin=2cm]{geometry}
\usepackage{graphicx,color}
\usepackage{subfigure}
\begin{document}
\maketitle
\begin{abstract}
In this paper, we consider a chaotic inflation model where the role of inflaton is played by the Higgs triplet in type II seesaw mechanism for generating the small masses of left-handed neutrinos. Leptogenesis could happen after inflation. This model is constructed without introducing supersymmetry (SUSY).
\end{abstract}
\footnoterule{\small $^1$chianshu@phys.sinica.edu.tw, $^2$cmlin@phys.nthu.edu.tw}
\section{Introduction}
Inflation (for the general review, \cite{textbook}) is perhaps the most popular model for the very early universe. It solves many problems of the hot big bang model and could provide the seeds of structure formation from quantum fluctuations in an inflating background. However, a big question for inflation is what is the inflaton field and how does it connect to particle physics. Without knowing this, we do not even know how to reheat the universe, hence recover the conventional hot big bang.

Among the many inflation models, chaotic inflation may be the most successful model for dealing with the initial condition of inflation\footnote{For an excellent review of chaotic inflation, see \cite{Linde:2007fr}.}. Inflation started immediately from Planck scale when the baby universe was created from quantum gravity. There is no need for a thermal equilibrium state before inflation in order to start inflation from phase transition. In addition, chaotic inflation predict a self reproducing process of the universe (or multiverse) which is called eternal inflation. It is almost becoming a common sense that we cannot have a model of inflation in the framework of standard model (SM)\footnote{However, if we consider non-minimal coupling to gravity, standard model Higgs could be the inflaton \cite{Bezrukov:2007ep}.} and we have to go beyond it. The most popular approach may be supersymmetry (SUSY). Chaotic inflation can be builded in SUSY by using right-handed sneutrino as the inflaton \cite{Murayama:1992ua, Murayama:1993xu}. However currently the real experimental evidences that we should go beyond the standard model is from neutrino oscillation which strongly indicates that neutrino has a small mass. In order to explain neutrino mass, seesaw mechanism is introduced. There are basically three types of seesaw mechanism. In type II seesaw, a Higgs triplet is introduced. The triplet scalar field can be implemented naturally in several contexts of physics beyond the SM. For example, in the left-right symmetry electroweak theory~\cite{Pati:1974yy} the gauge symmetry $SU(2)_{L}\times SU(2)_{R}\times U(1)_{B-L}$ breaks to the SM symmetry due to a right-triplet $T_{R} = (1,3,2)$. In $SU(5)$ grand unified theory a triplet scalar consisted in the fundamental representation \textbf{5} which breaks the SM to $U(1)_{Q}$. The minimal littlest Higgs model~\cite{ArkaniHamed:2002qy} in which the triplet Higgs scalar arises from the breaking of global $SU(5)$ down to $SO(5)$ symmetry as one of the Goldstone bosones. In this paper, we show that Higgs triplet can play the role of inflaton for chaotic inflation.

This paper is organized as follows. In section \ref{2}, we discuss type II seesaw and the constraints relevant to our model. In section \ref{3}, we present the inflation model by using Higgs triplet as the inflaton. In section \ref{4}, we discuss leptogenesis happens after inflation. Section \ref{5} is our conclusion.

\section{Type II Seesaw Mechanism}\label{2}
The type II seesaw mechanism enlarges the Higgs sector [$H = (H^{+}, H^{0})^{T}$] of the standard model with an isospin triplet, $\Delta$, of complex $SU(2)_{L}$ scalar fields with hypercharge $Y = 2$~\cite{type-II}. A Majorana mass for the observed neutrinos can be generated by a gauge-invariant Yukawa interaction of the left-handed lepton doublets with the scalar $\Delta$ without the introduction of the heavy right-handed neutrinos, the Yukawa reads
\begin{eqnarray}\label{Yukawa}
L_{Y} = Y_{ij}L_{iL}^{T}Ci\tau_{2}\Delta L_{jL} + {\rm H.c.}
\end{eqnarray}
Where the Yukawa couplings $Y_{ij}$ is a $3\times3$ symmetric complex matrix, $L_{iL}$ is the left-handed lepton doublet with flavor index $i = e, \mu, \tau$, $C$ is the charge conjugation operator, and $\tau_{2}$ is the Pauli matrix. The matrix representation of the triplet can be written as
\begin{eqnarray}
\Delta = \left(\begin{array}{cc}\Delta^{+}/\sqrt{2} & \Delta^{++} \\\Delta^{0} & -\Delta^{+}/\sqrt{2}\end{array}\right),
\end{eqnarray}
and the most general scalar potential is given by
\begin{eqnarray}\label{potential}
V(H,\Delta) &=& - \mu^2_{H}H^{\dag}H + \frac{\lambda}{4}(H^{\dag}H)^2 + M^2_{\Delta}Tr(\Delta^{\dag}\Delta) \nonumber \\
&& + \lambda_{1}(H^{\dag}H)Tr(\Delta^{\dag}\Delta) + \lambda_{2}[Tr(\Delta^{\dag}\Delta)]^2 \nonumber \\
&& + \lambda_{3}Tr(\Delta^{\dag}\Delta)^2 + \lambda_{4}H^{\dag}\Delta\Delta^{\dag}H \nonumber \\
&& + (\mu H^{T}i\tau_{2}\Delta^{\dag}H + {\rm H.c.}).
\end{eqnarray}
Here $\mu^2_{H} > 0$ to ensure the spontaneous breaking pattern of the SM via $\langle H^{0} \rangle = v/\sqrt{2}$, and $M^2_{\Delta} (> 0)$ is the mass term of the triplet scalars. In the limit of $\mu \rightarrow 0$ the symmetry of the model is enhanced, which leads to spontaneous violation of lepton number for $M_{\Delta} > 0$. The resulting massless scalar (so-called majoron, J) will contribute to the invisible width of $Z$ boson, and it is phenomenologically unacceptable as was excluded at LEP. Hence the simultaneous presence of the Yukawa interaction in Eq.~(\ref{Yukawa}) and the trilinear term $\mu(H^{\dag}i\tau_{2}\Delta^{\dag}H)$ with dimensionful parameter $\mu$ in Eq.~(\ref{potential}) will explicitly break lepton number and eliminate the majoron. The $\mu$-term may arise from the vacuum expectation value (VEV) of a scalar singlet field~\cite{Schechter:1981cv} or in the scenario of extra dimension~\cite{Ma:2000wpa}. Therefore the breaking of lepton number associated is communicated to the lepton sector through the VEV of the triplet scalar $\langle \Delta \rangle = v_{\Delta}/\sqrt{2}$. One expects that the Majorana mass term of neutrinos will be proportional to $Y_{ij}\times v_{\Delta}$. In this paper we consider a heavy triplet scalar, $M^2_{\Delta} \gg v^2$, so we will neglect the contributions from the terms involving $\lambda_{i} (i = 1-4)$. The value of the triplet VEV, $v_{\Delta}$ can be calculated from the minimum condition of the potential $V$, the results are
\begin{eqnarray}\label{minimum}
-\mu^2_{H} + \frac{\lambda}{4}v^2 - \sqrt{2}\mu v_{\Delta} = 0 \quad {\rm and} \quad v_{\Delta} = \frac{\mu v^2}{\sqrt{2}M^2_{\Delta}}.
\end{eqnarray}
The neutrino mass matrix can be generated via the Eq.~(\ref{Yukawa})
\begin{eqnarray}\label{numass}
M_{\nu} = \sqrt{2}Y_{ij}v_{\Delta} = Y_{ij}\frac{\mu v^2}{M^2_{\Delta}},
\end{eqnarray}
which can be realized the seesaw structure if we take $\mu \approx M_{\Delta}$.

The upper bound on the triplet VEV $v_{\Delta}$ can be obtained from the effect on $\rho$-parameter ($\rho = M^2_{W}/M^2_{Z}\cos^2{\theta_{W}}$)~\cite{Gunion:1990dt},
\begin{eqnarray}
\rho = 1 + \delta\rho = \frac{v^2 + 2v^2_{\Delta}}{v^2 + 4v^2_{\Delta}},
\end{eqnarray}
which is predicted to be $1$ in the SM. The experimental limit~\cite{PDG} leads to $v_{\Delta} \le {\cal O}(1)$ GeV. On the other hand, the present absolute neutrino masses are constrained through the electron energy spectrum from the end-point in the nuclear beta decays (i.e. the tritium decay, $m_{\beta} = \sqrt{\sum_{i}|U_{ei}|^2M^2_{\nu_{i}}} < 2$ eV)~\cite{PDG,Otten:2008zz} and the cosmological observations, $\sum_{i}M_{\nu_{i}} < 0.58$ eV (95\% CL)~\cite{Komatsu:2010fb}. As a result we have a lower bound of $v_{\Delta} > 1$ eV if we take the perturbative criterion for Yukawa coupling $Y_{ij} \le {\cal O}(1)$ in Eq.~(\ref{numass}). Consequently we obtain the range by using Eq.~(\ref{minimum})
\begin{eqnarray}
1~{\rm eV} \lae \frac{\mu v^2}{M^2_{\Delta}} \lae 1~{\rm GeV}.
\end{eqnarray}
In the limit of $\mu \approx M_{\Delta}$ and $v \sim {\cal O}(100)$ GeV the mass of triplet scalar is bounded in the range of
\begin{eqnarray}\label{mass}
10^4~{\rm GeV} \lae M_{\Delta} \lae 10^{13}~{\rm GeV}.
\end{eqnarray}

\section{Higg Triplet as the inflaton}
\label{3}

We assume the expectation value of $\Delta$ plays the role of the inflaton field $\phi$ with $\langle \Delta \rangle \equiv \frac{\phi}{\sqrt{2}}$ during inflation. The potential energy of $\phi$ can be read off from Eq.~(\ref{potential}) as
\begin{equation}\label{inflation}
V(\phi)=\frac{1}{2}M^2_\Delta \phi^2 \left(+\frac{\lambda_3}{4}\phi^4 \right).
\end{equation}
 Due the the large expectation value of $\phi$ during inflation, the effective mass of $H$ becomes very large from the last term in Eq.~(\ref{potential}). Therefore the expectation value of $H$ is driven to zero and we can neglect the last term in Eq.~(\ref{potential}). The quartic term is assumed to be negligible and we will consider its possible role later. Hence during inflation we have
\begin{equation}
V(\phi)=\frac{1}{2}M^2_\Delta \phi^2.
\end{equation}
This potential is ideal for chaotic inflation.
The slow roll parameters $\eta, \epsilon$ are given by
\begin{equation}
\eta \equiv M_P^2 \frac{V^{\prime\prime}}{V}=\epsilon \equiv \frac{1}{2} M_P^2 \left( \frac{V^\prime}{V}\right)^2=\frac{2M_P}{\phi^2}.
\end{equation}
The number of e-folds is
\begin{equation}
N=\frac{1}{M_P^2}\int \frac{V}{V^\prime}d\phi \simeq \frac{\phi^2}{4M_P^2}.
\end{equation}
The CMB scale corresponds to $N=60$ which makes $\phi \simeq 15 M_P$. The spectrum is given by
\begin{equation}
P_R=\frac{1}{24 \pi^2 M_P^4}\frac{V}{\epsilon}.
\end{equation}
CMB observation requires $P_R \simeq (5 \times 10^{-5})^2$ which makes $M_\Delta \simeq 10^{13}$ GeV. This value is favored by seesaw mechanism and agrees with Eq.~(\ref{mass}). The situation is similar to sneutrino inflation \cite{Murayama:1992ua, Murayama:1993xu}, but no SUSY is required in our setup. The spectral index $n_s$ is given by
\begin{equation}
n_s = 1+2\eta-6 \epsilon \simeq 0.967.
\end{equation}
The tensor to scalar ratio is
\begin{equation}
r=16\epsilon=0.13.
\end{equation}
This may be detectable from analysis of B-mode polarization of CMB data from PLANCK satellite \cite{:2006uk, Komatsu:2009kd}, the ground-based detectors QUIET+PolarBeaR \cite{Hazumi:2008zz}, or KEK's future CMB satellite experiment, LiteBIRD \cite{Hazumi:2008zz, LiteBird}.

     For the last term in Eq.~(\ref{inflation}) to be negligible at $N=60$, we need to have $\lambda_3 \lae 10^{-13}$. For the case $\lambda_3 \simeq 10^{-13}$, we can actually have a successful chaotic inflation driven by the quartic term. In this case, $M_\Delta$ can be smaller than $10^{13}$ GeV. However, chaotic inflation driven by a quartic term is on the verge of being ruled out \cite{Komatsu:2010fb}\footnote{However, see \cite{Ramirez:2009zs}.}.

\section{Reheating and Leptogenesis}
\label{4}
There are several channels that the inflaton $\phi$ can decay into, such as $\phi\rightarrow \nu\nu, HH$, and $ZZ$, with the decay widths given by
\begin{eqnarray}
\Gamma_\phi(\nu_{i}\nu_{j}) &\approx& \frac{Y^2_{ij}}{8\pi(1 + \delta_{ij})}M_{\Delta}, \\
\Gamma_{\phi}(HH) &\approx& \frac{M^3_{\Delta}v^2_{\Delta}}{8\pi v^4}, \\
\Gamma_{\phi}(ZZ) &\approx& \frac{g^2m^2_{Z}v^2_{\Delta}}{4\pi M_{\Delta}\cos^2{\theta_{W}}v^2}.
\end{eqnarray}
Here we neglect the mixings between $H$ and $T$. For estimate, we may assume the total decay rate of the inflaton is $\Gamma_{\phi}(tot.) \sim 0.01M_\Delta$. Therefore the reheating temperature $T_{reh.}$ is
\begin{equation}\label{reheating}
T_{reh.} \simeq 0.1\sqrt{\Gamma_\phi M_P} \simeq 10^{13}\mbox{ GeV}.
\end{equation}
For an inflation model based on SUSY, this value may be too high to cause gravitino problem. However, since we do not impose SUSY, we do not have gravitino problem.

Now we discuss the baryon asymmetry of the Universe through leptogenesis via the triplet scalar decay. In what follows we consider the decay final states of 2-lepton and 2-scalars. Since the appearance of $\mu$-term in the potential (Eq.~(\ref{potential})) the lepton number is explicitly violated due to the coexistence of the decays $\Delta \rightarrow LL$ and $\Delta \rightarrow HH$. CP violation will occur if there are extra contributions to the neutrino masses, such as with additional heavy fermion singlets~\cite{Hambye:2003ka,Guo:2004mp} or with another Higgs triplets~\cite{Ma:1998dx, Hambye:2000ui}. We concentrate on the later case and hence
the scale of lepton number violation is the same as the mass of the triplet Higgs scalars.
We write down the terms which are relevant to the leptogenesis
\begin{figure}[t]
  \centering
    \includegraphics[width=0.45\textwidth]{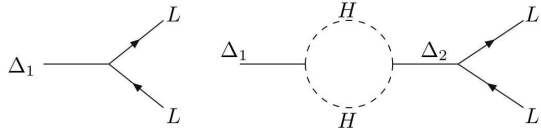}
  \caption{Lepton asymmetry in scalar triplet decays. }
  \label{fig:leptogenesis}
\end{figure}
\begin{eqnarray}
\label{20}
L &=& M^2_{\Delta_1}Tr(\Delta^{\dag}_{1}\Delta_{1}) + M^2_{\Delta_2}Tr(\Delta^{\dag}_{2}\Delta_{2}) + \{Y_{1ij}L_{iL}^{T}Ci\tau_{2}\Delta_{1}L_{jL} \nonumber \\
&+& Y_{2ij}L_{iL}^{T}Ci\tau_{2}\Delta_{2}L_{jL} + \mu_{1}H^{T}i\tau_{2}\Delta^{\dag}_{1}H + \mu_{2}H^{T}i\tau_{2}\Delta^{\dag}_{2}H + {\rm H.c.}  \}
\end{eqnarray}
The CP asymmetry $\epsilon_1$ is defined as
\begin{eqnarray}
\epsilon_1 &=& 2\frac{\Gamma(\Delta_1\rightarrow LL) - \Gamma(\bar{\Delta}_1\rightarrow \bar{L}\bar{L})}{\Gamma_{\Delta_1} + \Gamma_{\bar{\Delta}_1}}
\end{eqnarray}
with $2$ represents the processes violate lepton number by two units and $\Gamma_{\Delta_{1}}$ is the thermally averaged decay width of $\Delta_{1}$. CP asymmetry arises in the interferece of the tree with the one-loop self-energy diagrams as shown in Fig.~\ref{fig:leptogenesis} and will induce the off-diagonal mass matrix corrections. Here we assume $M_{\Delta_{1}} < M_{\Delta_{2}}$, so we can identify $\Delta_{2}$ as the inflaton $\phi$ in our scenario. Due to the large expectation value of $\Delta_2$ during inflation, the field value of $\Delta_1$ is driven to zero by a term $\sim |\Delta_1|^2 |\Delta_2|^2$, therefore it plays no role (such as a curvaton or two-field inflation) during inflation. After reheating when the temperature of the universe cooled down to below $M_{\Delta_{2}}$, most of $\Delta_{2}$ would decay away. However, the lepton asymmetry will be erased by the thermal equilibrium processes via the interactions of $\Delta_{1}$. So the asymmetry of the universe can only be generated by the subsequent decay of $\Delta_{1}$ at the temperature around $T\lae M_{\Delta_{1}}$. In order to have a successful leptogenesis, the $\mu$ terms in Eq.~(\ref{20}) has to exist during the reheating process\footnote{This means if the $\mu$ terms arise from spontaneous symmetry breaking of some scalar fields, those fields must already sit on their VEVs. We assume this is the case. It can be achieved if the masses of the symmetry breaking fields are larger than the Hubble parameter.}. We can interpret the leptogenesis as well as CP violation are created from the triplet scalar $\Delta_{1}$ "oscillate" into inflaton $\Delta_{2}$ and decay. A complete analysis of the leptogenesis with Higgs triplets is studied in the literature~\cite{Hambye:2003ka,Guo:2004mp,Ma:1998dx, Hambye:2000ui, Flanz:1996fb}. Here we consider the region where the mass square difference is much larger than the decay widths. The resulting lepton asymmetry of the decay is given by
\begin{eqnarray}
\epsilon_{1} &\approx& \frac{\rm{Im}[\mu_{1}\mu^*_{2}\sum_{k,l}(Y_{1kl}Y^*_{2kl})]}{8\pi^2(M^2_{\Delta_1} - M^2_{\Delta_2})}\Big(\frac{M_{\Delta_{1}}}{\Gamma_{\Delta_1}}\Big).
\end{eqnarray}
If we consider only $LL$ and $HH$ decay modes, {\it i.e.} $BR_{L} + BR_{H} =1$, one obtains an upper bound on this asymmetry
\begin{eqnarray}
\epsilon_{1} &\le& \frac{M_{\Delta_1}}{4\pi v^2}\sqrt{BR_{L}BR_{H}\sum_{i}m^2_{\nu_{i}}}.
\end{eqnarray}
As a result the asymmetry $\epsilon_{1}$ increases with larger $m_{\nu_{i}}$~\cite{Hambye:2003ka} unlike the canonical leptogenesis of the decaying fermion-singlet which is bounded by the absolute neutrino mass scale~\cite{Buchmuller:2003gz}. We define the parameter $K=\Gamma_{\Delta_{1}}/H(T=M_{\Delta_{1}})$ which is given by
\begin{eqnarray}\label{efficiency}
K \simeq \frac{10}{\sqrt{BR_{L}BR_{H}}}\left(\frac{|m_{\nu}|}{0.05 \rm{eV}}\right)
\end{eqnarray}
with $ H(T)|_{T=M_{\Delta_{1}}} = \sqrt{\frac{4\pi^3g_{*}}{45}}\frac{M_{\Delta_{1}}^2}{M_{P}}$ and $g_{*}\sim100$ is the effective number of massless particles. The Boltzmann equations read
\begin{eqnarray}
\frac{dY_{\Delta_{1}}}{dz} &=& -zK\left[\gamma_{D}(Y_{\Delta_{1}} - Y_{\Delta_{1}}^{eq}) + \gamma_{A}\frac{(Y^2_{\Delta_{1}} - Y_{\Delta_{1}}^{eq^2})}{Y_{\Delta_{1}}^{eq}}\right], \\
\frac{Y_{(\Delta_{1}-\bar{\Delta}_{1})}}{dz} &=& -zK\gamma_{D}\left[Y_{(\Delta_{1}-\bar{\Delta}_{1})} - \sum_{i}2BR_{i}\frac{Y_{\Delta_{1}^{eq}}}{Y_{i}^{eq}}Y_{i} \right], \\
\frac{Y_{L}}{dz} &=& 2zK\gamma_{D}\left[\epsilon_{1}(Y_{\Delta_{1}}-Y_{\Delta_{1}^{eq}}) + BR_{i}(Y_{(\Delta_{1} - \bar{\Delta}_{1})} - 2\frac{Y_{\Delta_{1}}^{eq}}{Y_{i}^{eq}}Y_{i}) \right],
\end{eqnarray}
where $z=M_{\Delta_{1}}/T$ and $Y$'s are the number densities per entropy density $s$ as defined by $Y_{\Delta_{1}} = n_{\Delta_{1}}/s$, $Y_{(\Delta_{1}-\bar{\Delta}_{1})} = (n_{\Delta_{1}} - n_{\bar{\Delta}_{1}})/s$, and $Y_{L} = (n_{L}-n_{\bar{L}})/s$. $\gamma_{D}$ and $\gamma_{A}$ are the quantities of decay, inverse-decay, and annihilation processes that affect the abundance of $\Delta_{1}$ and asymmetry, they are given by
\begin{eqnarray}
\gamma_{D} = \frac{K_{1}(z)}{K_{2}(z)} \quad {\rm and} \quad \gamma_{A} = \frac{T}{32\pi^4}\int_{4M^2_{\Delta_{1}}}^{\infty}ds(s-4M^2_{\Delta_{1}})\sigma_{A}\sqrt{s}K_{1}(\sqrt{s}/T)
\end{eqnarray}
with $K_{1,2}$ are the first and second kind modified Bessel functions and $\sigma_{A} \approx g^4/(\pi\sqrt{s(s-4M^2_{\Delta_{1}})})$ is the annihilation cross section due to the gauge interactions. Here we neglect the scattering of $LL \leftrightarrow HH$ as the assumption of small $\lambda$'s in the potential. In the case of $K > 1$ (see Eqs.~(\ref{reheating}),(\ref{efficiency})) the baryon asymmetry can be approximated by~\cite{kolb}
\begin{eqnarray}
\frac{n_{B}}{s} \sim 0.3\times10^{-2}\epsilon_{1}\times\left[K(\ln{K})^{0.6}\right]^{-1}.
\end{eqnarray}
For $M_{\Delta_{2}} = 4\times10^{13}$ GeV, $M_{\Delta_{1}} = 10^{13}$ GeV, $\mu_{1} = \mu_{2} = 10^{12}$ GeV, $1/\sqrt{BR_{L}BR_{H}} = 0.5$, and $m_{\nu} = 0.1$ eV, we have $n_{B}/s \approx 6\times10^{-10}$ as observed. We note that the result is insensitive to the mass of $M_{\Delta_{1}}$.
\section{Conclusions}
\label{5}
In this paper, we show that it is possible to have chaotic inflation by using the Higgs triplet in type II seesaw model as the inflaton. The required inflaton mass matches the mass we need for seesaw mechanism. It is also shown that leptogenesis could follow after the end of inflation in our setup. The model can be embedded in grand unified theory, left-right symmetry, little Higgs models, or supersymmetry. It is interesting to note that in the supersymmetric limit the masses of $\Delta_{1}$ and $\Delta_{2}$ are degenerate, then the soft supersymmetry breaking terms may provide the necessary mass splitting and CP violation for the resonant leptogenesis. However, in this case we may have to worry the gravitino problem as considering the reheating temperature produced after inflation is estimated to be roughly $10^{13}$ GeV. One may worry about that quantum corrections may destroy the flatness of the scalar potential when $\phi>M_P$, however since slow roll conditions ensure an approximate shift symmetry to the potential, the quantum corrections is logarithmic and negligible \cite{Kaloper:2008fb}.

\section*{Acknowledgement}
This work was supported in part by the
NSC under grant No. NSC 99-2811-M-007-068, by the NCTS, and by the
Boost Program of NTHU. CML would like to thank KEK for hospitality.



\begin{thebibliography}{99}

\bibitem{textbook}
 A.~D.~Linde,
 ``Particle Physics and Inflationary Cosmology,''
  Chur, Switzerland: Harwood (1990) 362 p;
 D.~H.~Lyth and A.~Riotto,
 Phys.\ Rept.\  {\bf 314}, 1 (1999);
 A.~R.~Liddle and D.~H.~Lyth
 ``Cosmological inflation and large-scale structure,''
{\it   Cambridge, UK: Univ. Pr. (2000) 400 p} ;
V.~Mukhanov,
 ``Physical foundations of cosmology,''
{\it  Cambridge, UK: Univ. Pr. (2005) 421 p};
 S.~Weinberg,
 ``Cosmology,''
{\it  Oxford, UK: Oxford Univ. Pr. (2008) 593 p};
 D.~H.~Lyth and A.~R.~Liddle,
 ``The primordial density perturbation: Cosmology, inflation and the origin of
 structure,''
{\it  Cambridge, UK: Cambridge Univ. Pr. (2009) 497 p};
 A.~Mazumdar and J.~Rocher,
 arXiv:1001.0993 [hep-ph].

\bibitem{Murayama:1992ua}
  H.~Murayama, H.~Suzuki, T.~Yanagida and J.~Yokoyama,
  Phys.\ Rev.\ Lett.\  {\bf 70}, 1912 (1993).

\bibitem{Murayama:1993xu}
  H.~Murayama, H.~Suzuki, T.~Yanagida and J.~Yokoyama,
  Phys.\ Rev.\  D {\bf 50}, 2356 (1994)
  [arXiv:hep-ph/9311326].

\bibitem{Linde:2007fr}
  A.~D.~Linde,
  Lect.\ Notes Phys.\  {\bf 738}, 1 (2008)
  [arXiv:0705.0164 [hep-th]].

\bibitem{Bezrukov:2007ep}
  F.~L.~Bezrukov and M.~Shaposhnikov,
  Phys.\ Lett.\  B {\bf 659}, 703 (2008)
  [arXiv:0710.3755 [hep-th]].

\bibitem{Pati:1974yy}
  J.~C.~Pati and A.~Salam,
  Phys.\ Rev.\  D {\bf 10}, 275 (1974)
  [Erratum-ibid.\  D {\bf 11}, 703 (1975)]; R.~N.~Mohapatra and J.~C.~Pati,
  Phys.\ Rev.\  D {\bf 11} 566, 2558 (1975); G.~Senjanovic and R.~N.~Mohapatra,
  Phys.\ Rev.\  D {\bf 12}, 1502 (1975); R.~E.~Marshak and R.~N.~Mohapatra,
  Phys.\ Lett.\  B {\bf 91}, 222 (1980).

\bibitem{ArkaniHamed:2002qy}
  N.~Arkani-Hamed, A.~G.~Cohen, E.~Katz and A.~E.~Nelson,
  JHEP {\bf 0207}, 034 (2002)
  [arXiv:hep-ph/0206021].


\bibitem{type-II} W.~Konetschny and W.~Kummer,
  Phys.\ Lett.\  B {\bf 70}, 433 (1977); J.~Schechter and J.~W.~F.~Valle,
  Phys.\ Rev.\  D {\bf 22}, 2227 (1980); T.~P.~Cheng and L.~F.~Li,
  Phys.\ Rev.\  D {\bf 22}, 2860 (1980); M.~Magg and C.~Wetterich,
  Phys.\ Lett.\  B {\bf 94}, 61 (1980); G.~Lazarides, Q.~Shafi and C.~Wetterich,
  Nucl.\ Phys.\  B {\bf 181}, 287 (1981); G.~B.~Gelmini and M.~Roncadelli,
  Phys.\ Lett.\  B {\bf 99}, 411 (1981); R.~N.~Mohapatra and G.~Senjanovic,
  Phys.\ Rev.\  D {\bf 23}, 165 (1981).

\bibitem{Schechter:1981cv}
  J.~Schechter and J.~W.~F.~Valle,
  Phys.\ Rev.\  D {\bf 25}, 774 (1982); M.~A.~Diaz, M.~A.~Garcia-Jareno, D.~A.~Restrepo and J.~W.~F.~Valle,
  Nucl.\ Phys.\  B {\bf 527}, 44 (1998)
  [arXiv:hep-ph/9803362].

\bibitem{Ma:2000wpa}
  E.~Ma, M.~Raidal and U.~Sarkar,
  Phys.\ Rev.\ Lett.\  {\bf 85}, 3769 (2000)
  [arXiv:hep-ph/0006046]; E.~Ma, M.~Raidal and U.~Sarkar,
  Nucl.\ Phys.\  B {\bf 615}, 313 (2001)
  [arXiv:hep-ph/0012101].

\bibitem{Gunion:1990dt}
  J.~F.~Gunion, R.~Vega and J.~Wudka,
  Phys.\ Rev.\  D {\bf 43}, 2322 (1991).

\bibitem{PDG} C. Amsler {\it et al.} (Particle Data Group), Physics Letters B667,1 (2008) and 2009 partial update for the 2010 edition.

\bibitem{Otten:2008zz}
  E.~W.~Otten and C.~Weinheimer,
  Rept.\ Prog.\ Phys.\  {\bf 71} (2008) 086201.


\bibitem{:2006uk}
    [Planck Collaboration],
  arXiv:astro-ph/0604069.


\bibitem{Komatsu:2009kd}
  E.~Komatsu {\it et al.},
  arXiv:0902.4759 [astro-ph.CO].

\bibitem{Hazumi:2008zz}
  M.~Hazumi,
  AIP Conf.\ Proc.\  {\bf 1040}, 78 (2008).

\bibitem{LiteBird}
http://cmbpol.kek.jp/litebird/

\bibitem{Komatsu:2010fb}
  E.~Komatsu {\it et al.},
  arXiv:1001.4538 [astro-ph.CO].

\bibitem{Ramirez:2009zs}
  E.~Ramirez and D.~J.~Schwarz,
  Phys.\ Rev.\  D {\bf 80}, 023525 (2009)
  [arXiv:0903.3543 [astro-ph.CO]].


\bibitem{Hambye:2003ka}
  T.~Hambye and G.~Senjanovic,
  Phys.\ Lett.\  B {\bf 582}, 73 (2004)
  [arXiv:hep-ph/0307237].

\bibitem{Guo:2004mp}
  P.~J.~O'Donnell and U.~Sarkar,
  Phys.\ Rev.\  D {\bf 49}, 2118 (1994)
  [arXiv:hep-ph/9307279]; W.~l.~Guo,
  Phys.\ Rev.\  D {\bf 70}, 053009 (2004)
  [arXiv:hep-ph/0406268]; S.~Antusch and S.~F.~King,
  Phys.\ Lett.\  B {\bf 597}, 199 (2004)
  [arXiv:hep-ph/0405093]; P.~H.~Gu, H.~Zhang and S.~Zhou,
  Phys.\ Rev.\  D {\bf 74}, 076002 (2006)
  [arXiv:hep-ph/0606302].

\bibitem{Ma:1998dx}
  E.~Ma and U.~Sarkar,
  Phys.\ Rev.\ Lett.\  {\bf 80}, 5716 (1998)
  [arXiv:hep-ph/9802445].

\bibitem{Hambye:2000ui}
  T.~Hambye, E.~Ma and U.~Sarkar,
  Nucl.\ Phys.\  B {\bf 602}, 23 (2001)
  [arXiv:hep-ph/0011192]; G.~D'Ambrosio, T.~Hambye, A.~Hektor, M.~Raidal and A.~Rossi,
  Phys.\ Lett.\  B {\bf 604}, 199 (2004)
  [arXiv:hep-ph/0407312]; E.~J.~Chun and S.~Scopel,
  Phys.\ Lett.\  B {\bf 636}, 278 (2006)
  [arXiv:hep-ph/0510170]; E.~J.~Chun and S.~Scopel,
  Phys.\ Rev.\  D {\bf 75}, 023508 (2007)
  [arXiv:hep-ph/0609259].

\bibitem{Flanz:1996fb}
  M.~Flanz, E.~A.~Paschos, U.~Sarkar and J.~Weiss,
  Phys.\ Lett.\  B {\bf 389}, 693 (1996)
  [arXiv:hep-ph/9607310] and
  Phys.\ Lett.\  B {\bf 345}, 248 (1995)
  [Erratum-ibid.\  B {\bf 382}, 447 (1996)]
  [arXiv:hep-ph/9411366]; A.~S.~Joshipura, E.~A.~Paschos and W.~Rodejohann,
  Nucl.\ Phys.\  B {\bf 611}, 227 (2001)
  [arXiv:hep-ph/0104228]; S.~Antusch and S.~F.~King,
  JHEP {\bf 0601}, 117 (2006)
  [arXiv:hep-ph/0507333]; T.~Hambye, M.~Raidal and A.~Strumia,
  Phys.\ Lett.\  B {\bf 632}, 667 (2006)
  [arXiv:hep-ph/0510008];  S.~Antusch,
  Phys.\ Rev.\  D {\bf 76}, 023512 (2007)
  [arXiv:0704.1591 [hep-ph]]; W.~Chao, S.~Luo and Z.~z.~Xing,
  Phys.\ Lett.\  B {\bf 659}, 281 (2008)
  [arXiv:0704.3838 [hep-ph]]; T.~Hallgren, T.~Konstandin and T.~Ohlsson,
  JCAP {\bf 0801}, 014 (2008)
  [arXiv:0710.2408 [hep-ph]].

\bibitem{Buchmuller:2003gz}
  W.~Buchmuller, P.~Di Bari and M.~Plumacher,
  Nucl.\ Phys.\  B {\bf 665}, 445 (2003)
  [arXiv:hep-ph/0302092].

\bibitem{kolb}
E.~W.~Kolb and M.~S.~Turner, {\it The Early Universe} (Addison-Wesley, Reading, MA, 1990).

\bibitem{Kaloper:2008fb}
  N.~Kaloper and L.~Sorbo,
  Phys.\ Rev.\ Lett.\  {\bf 102}, 121301 (2009)
  [arXiv:0811.1989 [hep-th]].


\end{thebibliography}
\end{document}